\documentclass[epsfig,preprint,showpacs,floatfix]{revtex4} 
\input{epsf.tex}
\usepackage{amssymb,epsfig}

\newcommand{\One}{1\kern-4.5pt1}
\newcommand{\lapprox}{\raisebox{-0.5ex}{$\ 
\stackrel{\textstyle<}{\textstyle\sim}\ $}}

\begin{document}
\renewcommand{\topfraction}{.95}
\renewcommand{\textfraction}{.1}

\title{The Shubnikov-de Haas effect in multiband quasi-two dimensional metals}
\author{I. O. Thomas$^1$, V. V. Kabanov$^{2}$ and A. S. Alexandrov$^1$}

\affiliation{$^1$Department of Physics, Loughborough University,
Loughborough, United Kingdom, LE11 3TU\\ $^{2}$Josef Stefan Institute 1001,
Ljubljana, Slovenia}

\begin{abstract}
We analyze the behavior of the longitudinal conductivity $\sigma_{zz}$ in a field perpendicular to the highly conducting plane of a quasi-2D {\em multiband} metal in the case of closed system and open systems where $\tau$ is fixed, $\tau$ oscillates due to intra-band scattering and $\tau$ oscillates due to inter-band scattering.  In all but one case, we find that there are also mixing frequencies present  -- however they exhibit different qualitative behaviors, as befits their different origins, and in the case of inter-band scattering in an opensystem, may in fact be absent in the dHvA oscillations of that system.
 
\end{abstract}


\maketitle
\section{Introduction}

The use of magnetic quantum oscillations \cite{Shoenberg} as a means of analyzing the Fermi surfaces of low-dimensional materials has received interest in recent years, particularly with respect to organic compounds such as the metallic varieties of the charge transfer salts BEDT-TFF (see reviews by Singleton \cite{Singleton} and Kartsovnik \cite{Kartsovnik}).  One interesting feature of such systems is that the magnetic quantum oscillations can render measurable an apparent difference between the canonical (closed) and grand canonical (open) ensembles (in which the density of electrons $n_e$ or the chemical potential $\mu$ are fixed, respectively); in the former case the chemical potential $\mu$ will oscillate, and this will have a noticeable effect on the behavior of the magnetic oscillations of a two- or quasi-two-dimensional system.

In two dimensional multi-band metals, frequency mixing in the de Haas-van Alphen effect resulting from chemical potential magneto-oscillations has been predicted \cite{alebra}, studied analytically and numerically \cite{alebra,nak,alebra2,cham,comment,taut,alebra3,fort} and observed experimentally \cite{she,other,Kartsovnik} in closed systems.  The theoretical description  has been generalized to the more realistic quasi-two-dimensional (quasi-2D) case \cite{alebra4}.  Recently, it was proposed that finite amplitudes due to mixing between extremal cross-sections of the warped Fermi surface of a {\em single} band closed quasi-2D system might also be observed \cite{AlexKab}; unlike the previous examples, this particular mixing effect cannot be replicated by magnetic breakdown \cite{Kartsovnik,MB}.

Attention has recently turned to the Shubnikov-De Haas (SdH) oscillations of the longitudinal conductivity of quasi-2D single-band metals, with quantum transport theory being used by Champel and Mineev \cite{min} to examine the ultra-high field, 2D limit, and by Grigoriev \cite{grig} to examine the intermediate (quasi-2D) field limit.  While much regarding the detailed physics of the former case remains obscure \cite{min1}, analysis of the latter case while taking into account scattering in the self-consistent Born approximation with $t\gg T_D$, where $T_D$ is the Dingle temperature and $t$ the interplane transfer integral, has resulted in a proposed explanation of a few oddities in the observed oscillations -- namely, the existence of slow oscillations and the field dependent phase shifts of the beats of the conductivity.

In what follows, we examine the behavior of the longitudinal conductivity $\sigma_{zz}$ in a field perpendicular to the highly conducting plane of a quasi-2D {\em multiband} metal.  We begin by deriving an expression for the conductivity of a two-band metal in the abscence of any scattering or chemical potential oscillations in Section \ref{sec:fixedtau}.  We then (Section \ref{sec:SEInter}) derive an expression for the self-energy of a system in the Born approximation where there exists scattering between two bands, which we will have need of in some of our following calculations.

In the section following that, we calculate the effects of chemical potential oscillations and different forms of scattering on the system.  Firstly, in Section \ref{sec:omunoscat} we consider a closed system with a fixed relaxation time $\tau$, then in Section \ref{sec:GPIntra} an open system with oscillations in $\tau$ arising from intra-band scattering (this is merely a two-band generalisation of Grigoriev's \cite{grig} result), in Section \ref{sec:CPIntra} a closed system with oscillations in $\tau$ arising from intra-band scattering, in Section \ref{sec:GPcond} an open system with oscillations in $\tau$ arising from inter-band scattering, and finally in Section \ref{sec:CPcond} a closed system  with oscillations in $\tau$ arising from inter-band scattering .  In all cases other than the second we discover some measure of frequency mixing, though the precise behaviour of such mixing varies according to its origin.  Interestingly, as argued at the beginning of Section \ref{sec:calc}, one would probably not expect frequency mixing due to inter-band scattering to be very significant in the dHvA effect; that it is present in open systems in the SdH effect seems to indicate that one distinction between open and closed low-dimensional thermodynamic systems is slightly blurred in the case of conductivity.

\section{SdH effect with fixed relaxation rate}

\subsection{General expression for longitudinal conductivity}
\label{sec:fixedtau}

A quasi-2D metal in a longitudinal magnetic field has the following energy spectrum \cite{Singleton,Kartsovnik}:
\begin{equation}
\epsilon_\alpha (n,k_z)=\omega_\alpha(n+1/2) - 2t\cos(k_z d),
\end{equation}

\noindent where $\alpha$ labels the band, $d$ is the distance between layers, $n$ labels the Landau level, $k_z$ is the momentum in the perpendicular direction and $\omega_\alpha=eB/m_\alpha$ is the cyclotron frequency, $m_\alpha$ being the effective mass for that band, and $\hbar=c=k_B=1$.

We may calculate the conductivity from the Kubo formula \cite{Kubo}
\begin{equation}
\sigma_{zz}=\frac{\pi e^2}{V}\int_{-\infty}^{\infty} d\xi [-n_F'(\xi)]\overline{\mbox{Tr}\left[\delta(\xi-H)\hat v_z\delta(\xi-H)\hat v_z\right]},
\end{equation}

\noindent with $\hat v_z$ being the velocity operator for the $z$-direction, $n_F'$ being the derivative of the Fermi function, the trace being taken over all the single particle states $\beta=\{n,k_z,k_y,\alpha\}\equiv\{\gamma,\alpha\}$ and the spin, $V$ is the volume of the system, and the bar denotes the averaging over the random distribution of impurities in the sample.   We may, if the impurities are point-like, their concentration is low and $t\gg T_D$, neglect the vertex corrections to the average and replace $\overline{\delta(\xi-\epsilon_\beta)}$ with $\Im G^R(\epsilon_\beta,\xi)/\pi$, where 
\begin{equation}
G^R(\epsilon_\beta,\xi)= \frac{1}{\xi-\epsilon_\beta-\Sigma_\beta(\xi)},\label{eqn:GR}
\end{equation}

\noindent is the retarded Green's function and $\Sigma_\beta(\xi)=L_\beta(\xi)-i\Delta_\beta(\xi)$ is the retarded self-energy ($L_\beta(\xi)$ and $\Delta_\beta(\xi)$ being real), to obtain
\begin{equation}
\sigma_{zz}= \frac{2e^2}{V\pi}\int d\xi[-n_F'(\xi)]\sum_\alpha \sum_\gamma v_{z\alpha,\,\gamma}^2[\Im G^R(\epsilon_{\alpha,\gamma},\xi)]^2.
\end{equation}

This is simply the initial conductivity formula of references \cite{grig,min}, only now generalized so that it sums over multiple bands.  Restricting our interest to the case of two bands with different masses $m_1$ and $m_2$, and since $t\gg T_D$,  we may  use the approximation $\Sigma_\beta(\xi)\approx\Sigma(\xi)$, making the calculation analytically tractable). Then  we may generalize the single-band result \cite{grig,min} as
\begin{eqnarray}
\sigma_{zz}&=&\frac {e^2N t d^2}{\pi}\int {d\xi}[-n_F'(\xi)]\sum_\alpha\sum_{k=-\infty}^{\infty}\frac{(-1)^k}{k}J_1\left(\frac{4\pi k t}{\omega_\alpha}\right)\nonumber\\
&&\times \exp\left(\frac{2\pi i k \xi^*}{\omega_\alpha}\right)\left(\frac{1}{\Delta(\xi)} + \frac{2\pi|k|}{\omega_\alpha}\right)R_{\alpha}(k,\xi). \label{eqn:startpnt}
\end{eqnarray}

Here, $J_1(x)$ is the first-order Bessel function, $\xi^*=\xi-\tilde L(\xi)$, $\tilde L(\xi)$ being the oscillating portion of $L(\xi)$, $N=eB/2d\pi$ and 
\begin{equation}
R_{\alpha}(k,\xi)=\exp\left(\frac{-2\pi\Delta(\xi)|k|}{\omega_\alpha}\right).\end{equation}

\noindent One should use the expansion $J_1(kx)/k=x/2$ for the $k=0$ harmonic.

To begin with, let us assume that the quantum oscillations in the self-energy are suppressed, with the result that we can ignore the oscillatory contribution to $\xi^*$ and $\Delta(\xi)=(2\tau)^{-1}=\pi T_D$, where $\tau$ is the mean scattering time and $T_D$ is the associated Dingle temperature. This is justified in the presence of a field and size independent reservoir of states \cite{alebra4,min}, for example, which suppresses both chemical potential oscillations and any oscillations in the self-energy.  Having made this assumption, we integrate over $\xi$ (using the delta-function-like behavior of $n_F'(\xi)$ near the Fermi energy at small $T$ to obtain the first term, and $\int_{-\infty}^{\infty}\cos(ax)\cosh^{-2}(x)dx= a\pi/\sinh(a\pi/2)$ to obtain the damping factor for the oscillatory terms) and so acquire:
\begin{equation}
\sigma_{zz}=
\sigma\left[1+\sum_\alpha \frac{m_\alpha}{M}\sum_{k=1}^{\infty}\frac{(-R_{\alpha})^k}{k}J_1\left(\frac{4\pi k t}{\omega_\alpha}\right)\left(\frac{\omega_\alpha}{\pi t}+\frac{2\pi k T_D}{t}\right)\cos\left(\frac{2\pi k\mu}{\omega_\alpha}\right)\Theta\left(\frac{2\pi^2kT}{\omega_\alpha}\right)\right],
\end{equation}

\noindent where $\sigma=(e^2d t^2 M)/(\pi^2 T_D)$, $M=m_1+m_2$, $R_{\alpha}=\exp(-2\pi^2T_D/\omega_\alpha)$ is the Dingle reduction factor, and $\Theta(y)=y/\sinh(y)$  is the usual Lifshitz-Kosevitch reduction factor. If $4\pi t>\omega_\alpha$ we can use the asymptotic
\begin{equation}
J_1\left(\frac{4\pi k t}{\omega_\alpha}\right)\approx \sqrt{\frac{\omega_\alpha}{2k\pi^2t}}\sin\left(\frac{4\pi k t}{\omega_\alpha}-\frac{\pi}{4}\right),
\end{equation}

\noindent and so obtain
\begin{equation}
\sigma_{zz}=\sigma\left[1+M^{-1}\sum_\alpha m_\alpha \sum_{k=1}^{\infty}A_k^{\alpha}\cos\left(\frac{2\pi k\mu}{\omega_\alpha}\right)\Theta\left(\frac{2\pi^2kT}{\omega_\alpha}\right)\right]\label{eqn:ftaufmu},
\end{equation}

\noindent where
\begin{equation}
A_k^{\alpha}=\frac{(-R_{\alpha})^k}{k}\sqrt{\frac{\omega_\alpha}{2k\pi^2t}}\left(\frac{\omega_\alpha}{\pi t}+\frac{2\pi k T_D}{t}\right)\sin\left(\frac{4\pi k t}{\omega_\alpha}-\frac{\pi}{4}\right).
\end{equation}

We can see then that the conductivity oscillates with two fundamental periods, one corresponding to the first band and one to the second.  In the quasi-3D limit one will observe the splitting of each fundamental in the Fourier Transform of the conductivity into a pair of peaks (a low-frequency `neck' peak and a high-frequency `belly' peak), which is evidence of the slight warping of the Fermi-surface such that there are two extremal orbits present\cite{Shoenberg}.


\subsection{Self-energy with interband scattering}
\label{sec:SEInter}

If scattering between bands is possible, then the self energy of the particles in any given band will contain contributions from all other bands in addition to that due to intra-band scattering.  In that case, in order to obtain the conductivity in the self-consistent Born approximation in the two band case, we must include the impurity diagrams from both bands in $\Sigma_\beta(\xi)$ \cite{Mahan}:
\begin{eqnarray}
\Sigma^R(\xi)&=& W\sum_\alpha\sum_\gamma G^R(\epsilon_{\alpha,\gamma},\xi),\label{eqn:symmapp}
\end{eqnarray}

\noindent where $W$ is the square of the scattering amplitude, which is proportional to the impurity concentration and set equal to the same constant in every band for simplicity.  It also includes a factor of 2 from the summation over the spin.
Following reference \cite{grig} we find that
\begin{eqnarray}
\sum_\gamma G^R(\epsilon_{\alpha,\gamma},\xi)&=&-\frac{Vm_\alpha}{2\pi d}\left[A(\xi)-2\pi\sum_{k=1}^\infty (-1)^k J_0\left(\frac{4\pi kt}{\omega_\alpha}\right)\sin\left(\frac{2\pi k\xi^*}{\omega_\alpha}\right)R_{\alpha}(k,\xi)\right.\nonumber\\
&&+\left.i\pi\left(1+2\sum_{k=1}^{\infty}(-1)^k J_0\left(\frac{4\pi kt}{\omega_\alpha}\right)\cos\left(\frac{2\pi k\xi^*}{\omega_\alpha}\right)R_{\alpha}(k,\xi)\right)\right].
\end{eqnarray}

From this, we can see that there is an oscillatory contribution from the real part of the self-energy that must be taken into account in our calculations, and a slowly varying part $A(\xi)$ which may be ignored.  Using our symmetric approximation (equation (\ref{eqn:symmapp})) and setting $WVM/2\pi d=\pi T_D$ (as the average Dingle temperature is simply related to the average value of the imaginary part of the self-energy due to scattering)  we may write the imaginary and oscillating real parts of the self energy as follows:
\begin{equation}
\Delta(\xi)=\pi T_D\left[1+2M^{-1}\sum_\alpha m_\alpha \sum_{k=1}^{\infty}(-1)^k J_0\left(\frac{4\pi kt}{\omega_\alpha}\right)\cos\left(\frac{2\pi k\xi^*}{\omega_\alpha}\right)R_{\alpha}(k,\xi)\right],\label{eqn:Deltasc}
\end{equation}
\begin{equation}
\tilde L(\xi)=2\pi T_D M^{-1} \sum_\alpha m_\alpha\sum_{k=1}^{\infty}(-1)^k J_0\left(\frac{4\pi kt}{\omega_\alpha}\right)\sin\left(\frac{2\pi k\xi^*}{\omega_\alpha}\right)R_{\alpha}(k,\xi).
\end{equation}

\noindent Equation (\ref{eqn:Deltasc}) is a non-linear equation for $\Delta(\xi)$ which can be solved approximately in the strong damping limit $R_{\alpha}(k,\xi)\ll1$, as we shall see in section \ref{sec:GPcond}.

\section{The effects of scattering and oscillating $\mu$ on the conductivity}
\label{sec:calc}

In this section, we discuss the effects of chemical potential oscillations and of various kinds of scattering on the conductivity.  It should be noted that this analysis takes place in a region where there is strong damping of the amplitudes by the factor $R_\alpha$ -- we do not work in the region where the processes giving rise to mixing frequencies are at their strongest, and that given that the effects we are interested in are of the second order in $R_\alpha$, they may be quite small.  However, it has been noted (by comparison to numerical results) in the case of dHvA oscillations \cite{comment,AlexKab} that a reasonable level of accuracy may be maintained even if one allows the Dingle damping factor to tend towards unity, assuming that one is in the appropriate limit regarding the behaviour of $\mu$.  This might also be the case with respect to the SdH oscillations where mixing due to inter-band scattering is present.

Indeed, it would be suprising if our analysis of those systems were not even qualitiatively correct (with regards to the presence of frequency mixing at the very least) outside of the region where $R_\alpha\ll1$, since the amplitudes contain a denominator $\Delta(\xi)$ which becomes more important as it becomes smaller and $R_\alpha$ therefore approaches unity.  Interestingly, this would not be the case with regards to dHvA oscillations -- as can be seen from equation (68) of \cite{Wasserman},  $\Delta(\xi)$ enters only through the Dingle damping factor and the cosine function.  This entails that, as $R_\alpha$ approaches 1, the oscillating portions of $\Delta(\xi)$ become negligible -- it is doubtful, therefore, that one would be able to observe frequency mixing due to scattering in the dHvA oscillations of any but the dirtiest systems.

In cases where slow oscillations of the conductivity exist, it should be noted that it has been observed \cite{Kartgrig} that macroscopic inhomogeneities in the sample increase the damping of the fast oscillations over and above that expected by simple scattering; this can be modeled by replacing the Dingle temperature in the $R_\alpha$ factors of those oscillations with a new, larger Dingle temperature $T_D^*$ that will enhance the damping of their amplitudes. In order to simplify our discussion, though, we ignore this complication for now, though it should be taken into account in any physical measurement.

\subsection{Canonical Ensemble: oscillating $\mu$, fixed $\tau$}
\label{sec:omunoscat}

In closed systems, the particle density is fixed and we work in the Canonical Ensemble.  As a result of this, the chemical potential $\mu$ of the system may oscillate. In three dimensions these oscillations are negligible, however, as mentioned previously, in the case of multiband 2D and quasi-2D metals with two bands one discovers additional harmonics corresponding to frequency mixing between bands.  In this section, we discuss the effects of these oscillations on the conductivity of a system in which scattering effects are suppressed.  This situation is physical if the oscillations in $\tau$ are suppressed due to a large amplitude of scattering from quantised bands to the resevoir, which is consistent with the existence of a small enough resevoir density of states for the oscillations in $\mu$ to remain significant.

Following reference \cite{alebra4} we divide the chemical potential into an oscillating part $\tilde\mu$ and a non-oscillating part $\mu_0$:
\begin{equation}
\mu=\mu_0+\tilde\mu,
\end{equation}

\noindent where
\begin{equation}
\tilde\mu=-\frac{1}{\rho}\left(\frac{\partial \tilde\Omega}{\partial \mu}\right),
\end{equation}

\noindent with $\tilde\Omega$ being the oscillatory portion of the thermodynamic potential $\Omega$, $\rho=\rho_{bg}+\sum_\alpha \rho_\alpha$, $\rho_{bg}$ being the unquantized background density of states which we treat as being negligible, and $\rho_\alpha$ being the quantized density of states for the band $\alpha$.  Using formula (14) of reference \cite{alebra4} and the 2D density of states $\rho_\alpha=m_\alpha/\pi$, we acquire
\begin{equation}
\tilde\mu=-\sum_\alpha \sum_{k=1}^{\infty}B_k^{\alpha}\sin\left(\frac{2\pi k (\mu_0+\tilde\mu)}{\omega_\alpha}\right)\Theta\left(\frac{2\pi^2kT}{\omega_\alpha}\right),\label{eqn:mutilde}
\end{equation}

\noindent where
\begin{eqnarray}
B_k^\alpha&=& \frac{2 eB(R_{\alpha})^k}{\pi Mk}J_0\left(\frac{4\pi k t}{\omega_\alpha}\right) \nonumber \\
&\approx&\frac{2 eB (R_{\alpha})^k}{\pi M k}\sqrt{\frac{\omega_\alpha}{2k\pi^2t}}\cos\left(\frac{4\pi k t}{\omega_\alpha}-\frac{\pi}{4}\right),
\end{eqnarray}

\noindent which follows from the use of the asymptotic
\begin{equation}
J_0\left(\frac{4\pi k t}{\omega_\alpha}\right)\approx \sqrt{\frac{\omega_\alpha}{2k\pi^2t}}\cos\left(\frac{4\pi k t}{\omega_\alpha}-\frac{\pi}{4}\right).
\end{equation}

Let us assume that the system is strongly damped: $R_\alpha\ll1$, and we keep only the oscillating terms of second order in $R_\alpha$ or less. It follows that $\tilde\mu/\omega_\alpha\ll1$, so we may expand (\ref{eqn:ftaufmu}) in $\tilde\mu$ and approximate $\mu_0$ with $\mu$.  We then insert (\ref{eqn:mutilde}), again approximating the argument of the sine function as $\mu\approx\mu_0$, and so obtain
\begin{equation}
\sigma_{zz}=\sigma^{ord}_{zz}+\sigma^{mix}_{zz}\label{eqn:tmuosc}
\end{equation}

\noindent where $\sigma^{ord}_{zz}$ represents the unmixed portion of the conductivity and $\sigma^{mix}_{zz}$ is the mixed portion. $\sigma^{ord}_{zz}$ is given by:
\begin{eqnarray}
\frac{\sigma_{zz}^{ord}}{\sigma}&=&1+M^{-1}\sum_\alpha \left(m_\alpha A_1^\alpha \Theta\left(\frac{2\pi^2T}{\omega_\alpha}\right)\cos\left(\frac{2\pi \mu}{\omega_\alpha}\right) \right. \nonumber\\
&&+\left[m_\alpha A_2^\alpha\Theta\left(\frac{4\pi^2T}{\omega_\alpha}\right)+C_{\alpha\alpha}\Theta\left(\frac{2\pi^2T}{\omega_\alpha}\right)^2\right]\cos\left(\frac{4\pi\mu}{\omega_\alpha}\right)\nonumber\\
&&\left.-C_{\alpha\alpha}\Theta\left(\frac{2\pi^2T}{\omega_\alpha}\right)^2\right),
\end{eqnarray}

\noindent which shows that the oscillations in $\tilde\mu$ modify the second harmonics and also create additional slow oscillations due to the effect of the warping of the Fermi surface on the behaviour of the scattering \cite{Kartgrig}, as predicted for the single band case by Grigoriev \cite{grig}.  This slow oscillation is not seen in dHvA oscillations (such as those of reference \cite{AlexKab}, for example); in those cases, one sees only the fast oscillations whose frequencies are determined by $t$ and $\mu$ together, with no slow oscillations whose frequencies are determined by $t$ alone.

\begin{figure}[htb]
\begin{center}
\epsfig{file=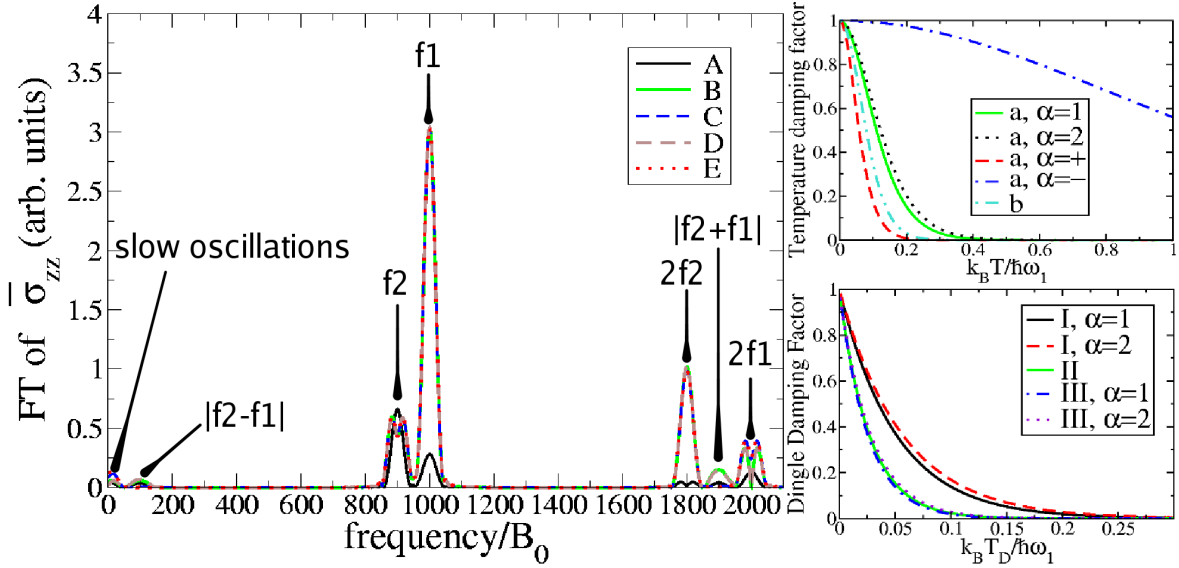, height=7.5cm}
\end{center}
\vspace{-.5cm}
\caption{\small A graph showing the Fourier transform of the SdH oscillations from $.9\le B_0/B\le.95$ where the frequencies are given in terms of $B/B_0$, alongside plots of how the damping factors evolve as their respective temperatures are increased. Due to the small size of the window, the slow oscillations are poorly resolved.The y axis has been rescaled by a factor of 100. The following parameters are used: $\mu=500t$, $(\omega_1/\omega_2)=(m_2/m_1)=.9$, $k_B T_D=.026t$ and $k_B T=.00005t$, setting our unit of measurement to be $B_0=m_1t/2\hbar e$,  which is around 41 Tesla if $t=0.01$eV and $m_1=m_e$.   The legend in the Fourier Transform plots  should be interpreted as follows: A -- closed system, fixed $\tau$; B -- open system, inter-band scattering; C -- closed system, intra-band scattering; D -- closed system, interband scattering; E -- open system, intraband scattering.  The legend in the Temperature Damping plots should be interpreted as: a -- value of $\Theta(2\pi^2 T/\omega_\alpha)$;  b -- value of $\Theta(2\pi^2 T/\omega_1)\Theta(2\pi^2 T/\omega_2)$.  The legend in the Dingle Damping plots should be interpreted as: I -- value of $R_\alpha$; II -- value of $R_1 R_{2}$; III -- value of $R_\alpha^2$.}
\label{fig:.9}
\end{figure}

The mixed component of the conductivity is given by
\begin{equation}
\frac{\sigma_{zz}^{mix}}{\sigma}= M^{-1}(C_{12}+C_{21})\Theta\left(\frac{2\pi^2T}{\omega_1}\right)\Theta\left(\frac{2\pi^2T}{\omega_2}\right)\left(\cos\left(\frac{2\pi\mu}{ \omega_+}\right)-\cos\left(\frac{2\pi\mu }{\omega_-}\right)\right),
\end{equation}

\noindent where 
\begin{equation}
\frac{1}{\omega_\pm}=\frac{1}{\omega_{2}}\pm\frac{1}{\omega_{1}},
\end{equation}

\noindent and it is this term which gives rise to the frequency mixing induced by the chemical potential oscillations.

The amplitudes induced by the chemical potential oscillation are given by:
\begin{equation}
C_{\alpha\alpha}=\frac{2eB m_\alpha a_{1,\,\alpha}}{M\pi^2 t}R_\alpha^2\sin\left(2\left[\frac{4\pi t}{\omega_\alpha} - \frac{\pi}{4}\right]\right),
\end{equation}

\noindent in the case where $\alpha=\alpha'$, and
\begin{equation}
C_{\alpha\alpha'}=\frac{2eB m_\alpha a_{1,\,\alpha}}{M\pi^2 t}\left(\frac{m_\alpha}{m_{\alpha'}}\right)^{1/2}R_\alpha R_{\alpha'}\left[\cos\left(\frac{4\pi t}{\omega_+}\right)\pm\sin\left(\frac{4\pi t}{\omega_-}\right)\right],
\end{equation}

\noindent in the mixed ($\alpha\ne\alpha'$) case, where
\begin{equation}
a_{r,\,\alpha}=\frac{1}{2}\left(\frac{\omega_\alpha}{r\pi t} + \frac{2\pi T_D}{t}\right),\label{eqn:adef}
\end{equation}

\noindent and we take the positive sign in front of the sine when $\alpha'=1$ and the negative when $\alpha'=2$.

One important feature of the mixing amplitudes in this case is that they are identical for both the additive and the subtractive mixing frequencies.  This does not generally hold, as we shall see in what follows.


\subsection{Grand Canonical Ensemble: Intra-band scattering}
\label{sec:GPIntra}

Working in the self-consistent Born approximation, let us assume that there is no scattering between the bands, and that we work in an open system described by the Grand Canonical Ensemble.  In this case the only diagrams contributing to the self energy of an electron in band $\alpha$ will be those corresponding to intra-band scattering, and there are no chemical potential oscillations that could also result in a mixing of oscillation frequencies.  This is the situation described by Grigoriev \cite{grig}.  Here we simply generalize it to the case of multiple bands.  Keeping all terms up to and including $O(R_{\alpha}^2)$, we obtain:
\begin{equation}
\sigma_{zz}=\sigma\left[1+M^{-1}\sum_\alpha m_\alpha \left(D_1^\alpha\Theta\left(\frac{2\pi^2 T}{\omega_\alpha}\right)\cos\left(\frac{2\pi\mu}{\omega_\alpha}\right)+D_2^\alpha\Theta\left(\frac{4\pi^2 T}{\omega_\alpha}\right)\cos\left(\frac{4\pi\mu}{\omega_\alpha}\right)+D_S^\alpha\right)\right], \label{eqn:scatt}
\end{equation}

where
\begin{equation}
D_1^\alpha=2\sqrt{\frac{\omega_\alpha}{2\pi^2t}\left(1+(a_{1,\,\alpha})^2\right)}R_{\alpha}\cos\left(\frac{4\pi t}{\omega_\alpha}-\frac{\pi}{4}+\phi_{1,\,\alpha}\right),\label{eqn:Ist1}
\end{equation}
\begin{eqnarray}
D_2^\alpha&=&\frac{4T_D}{t}\sqrt{1+(a_{1,\,\alpha})^2}(R_{\alpha})^2\cos\left(\frac{4\pi t}{\omega_\alpha}-\frac{\pi}{4}\right)\cos\left(\frac{4\pi t}{\omega_\alpha}-\frac{\pi}{4}+\phi_{1,\,\alpha}\right)\nonumber\\
&&+(R_{\alpha})^2\sqrt{\frac{\omega_\alpha}{\pi^2t}\left(1+(a_{2,\,\alpha})^2\right)}\cos\left(\frac{8\pi t}{\omega_\alpha}-\frac{\pi}{4}+\phi_{2,\,\alpha}\right)\nonumber +D_S^\alpha,
\end{eqnarray}
\begin{equation}
D_S^\alpha=\frac{\omega_\alpha (R_{\alpha})^2}{2\pi^2 t}\left[\sqrt{1+\left(\frac{\omega_\alpha}{2 \pi t}\right)^2}\cos\left(2\left[\frac{4\pi t}{\omega_\alpha}-\frac{\pi}{4}+\phi_{S\alpha}\right]\right) +1\right],
\end{equation}
\begin{equation}
\phi_{r,\,\alpha}=\arctan(a_{r,\,\alpha}),\,\phi_{S,\,\alpha}=\arctan\left(\frac{\omega_\alpha}{2 \pi t}\right),\label{eqn:angles}
\end{equation}

\noindent and $a_{r,\,\alpha}$ is given by equation (\ref{eqn:adef}).

\subsection{Canonical Ensemble: oscillating $\mu$, intra-band scattering}
\label{sec:CPIntra}

When the system is closed, we must take into account the effects of the oscillations in $\mu$ as well as that of scattering.  We can do this by applying the proceedure outlined in Section \ref{sec:omunoscat} to equation (\ref{eqn:scatt}), assuming that $R_\alpha \ll1$.  

\begin{figure}[htb]
\begin{center}
\epsfig{file=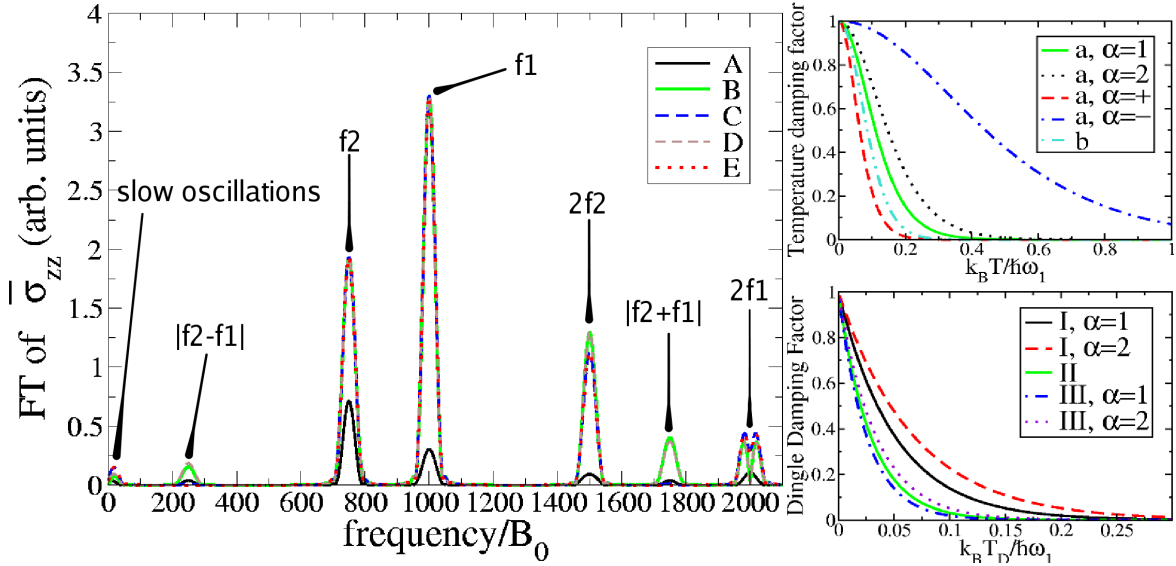, height=7.5cm}
\end{center}
\vspace{-.5cm}
\caption{\small A graph showing the Fourier transform of the SdH oscillations from $.9\le B_0/B\le.95$  with $(\omega_1/\omega_2)=(m_2/m_1)=.75$, alongside plots of how the damping factors evolve as their respective temperatures are increased. Due to the small size of the window, the slow oscillations are poorly resolved.The y axis has been rescaled by a factor of 100.  Except as otherwise noted, all parameters and legends are identical to those in Figure \ref{fig:.9}.}
\label{fig:.79}
\end{figure}

Having performed the expansion in terms of $\tilde\mu$, we may write our result as:
\begin{equation}
\sigma_{zz}=\sigma^{ord}_{zz}+\sigma^{mix}_{zz},
\end{equation}

\noindent where
\begin{eqnarray}
\frac{\sigma^{ord}_{zz}}{\sigma}&=&1+M^{-1}\sum_\alpha m_\alpha \left(D_1^\alpha\Theta\left(\frac{2\pi^2 T}{\omega_\alpha}\right)\cos\left(\frac{2\pi\mu}{\omega_\alpha}\right)\right.\nonumber\\
&&+\left[D_2^\alpha\Theta\left(\frac{4\pi^2 T}{\omega_\alpha}\right)
-\mathcal{C}_{\alpha\alpha}\Theta\left(\frac{2\pi^2 T}{\omega_\alpha}\right)^2
\right]\cos\left(\frac{4\pi\mu}{\omega_\alpha}\right)\nonumber\\
&&\left.+D_S^\alpha + \mathcal{C}_{\alpha\alpha}\Theta\left(\frac{2\pi^2 T}{\omega_\alpha}\right)^2\right),
\end{eqnarray}

\noindent and
\begin{equation}
\frac{\sigma^{mix}_{zz}}{\sigma}=M^{-1}\left(m_1\mathcal{C}_{12}+m_2\mathcal{C}_{21}\right)\Theta\left(\frac{2\pi^2 T}{\omega_1}\right)\Theta\left(\frac{2\pi^2 T}{\omega_2}\right)\left(\cos\left(\frac{2\pi\mu}{\omega_-}\right)-\cos\left(\frac{2\pi\mu}{\omega_+}\right)\right).
\end{equation}

The amplitudes due to the chemical potential oscillations are given by
\begin{equation}
\mathcal{C}_{\alpha\alpha}=\frac{eB R_\alpha^2}{M\pi^2t}\sqrt{1+(a_{r,\,\alpha})^2}\left[\cos(\phi_{1,\,\alpha})+\cos\left(2\left[\frac{4\pi t}{\omega_\alpha}-\frac{\pi}{4}+\frac{\phi_{1,\,\alpha}}{2}\right]\right)\right]
\end{equation}

\noindent when $\alpha=\alpha'$ and by the following when $\alpha\ne\alpha'$:
\begin{equation}
\mathcal{C}_{\alpha\alpha'}=\frac{eB R_\alpha R_{\alpha'}}{M\pi^2t}\left(\frac{m_\alpha}{m_{\alpha'}}\right)^{1/2}\sqrt{1+(a_{r,\,\alpha})^2}\left[\cos\left(\frac{4\pi t}{\omega_-}\pm\phi_{1,\,\alpha}\right)-\sin\left(\frac{4\pi t}{\omega_+}+\phi_{1,\,\alpha}\right)\right],
\end{equation}

\noindent where the argument of the cosine function contains a `$+$' if $\alpha=2$ and a `$-$' if $\alpha=1$.  

In this case, we find that we have two terms contributing to the slow oscillations, one of which is temperature dependent, and that (as one might expect from section \ref{sec:omunoscat}) we have additional frequencies due to the mixing of bands by the oscillations in the chemical potential.


\subsection{Grand Canonical Ensemble: inter-band scattering}
\label{sec:GPcond}

In order to proceed with our calculation of the effects of inter-band scattering on the behaviour of the system in the Grand Canonical Ensemble, we expand (\ref{eqn:Deltasc}) in the strong damping limit $R_\alpha(k,\xi)\ll 1$, where we make the approximation 
\begin{equation}
R_{\alpha}(1,\xi)\approx R_{\alpha}\left[1+\frac{4\pi^2T_D}{\omega_\alpha M}\sum_{\alpha'} m_{\alpha'} J_0\left(\frac{4\pi t}{\omega_{\alpha'}}\right)\cos\left(\frac{2\pi \xi}{\omega_{\alpha'}}\right)R_{\alpha'}\right],\,R_{\alpha}(2,\xi)\approx R_{\alpha}^2 \label{eqn:R1sub}
\end{equation}

\begin{figure}[htb]
\begin{center}
\epsfig{file=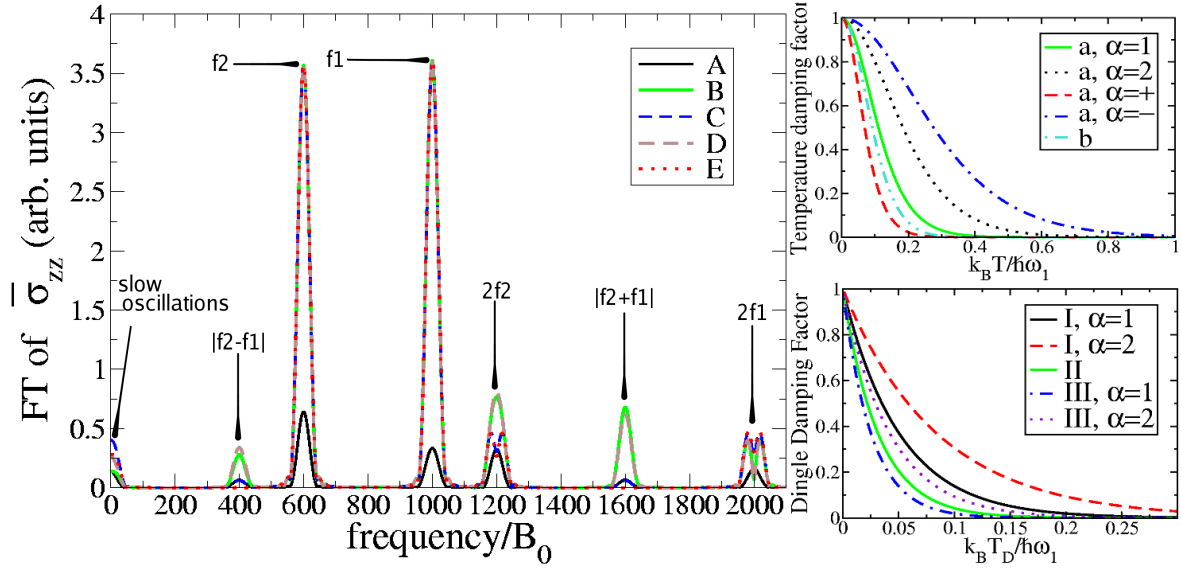, height=7.5cm}
\end{center}
\vspace{-.5cm}
\caption{\small A graph showing the Fourier transform of the SdH oscillations from $.9\le B_0/B\le.95$  with $(\omega_1/\omega_2)=(m_2/m_1)=.6$, alongside plots of how the damping factors evolve as their respective temperatures are increased. Due to the small size of the window, the slow oscillations are poorly resolved.The y axis has been rescaled by a factor of 100.  Except as otherwise noted, all parameters and legends are identical to those in Figure \ref{fig:.9}.}
\label{fig:.6}
\end{figure}

We can then expand out the cosine terms for small $\tilde L(\xi)$, and find that for $k=1$:
\begin{eqnarray}
\cos\left(\frac{2\pi \xi^*}{\omega_\alpha}\right)R_{\alpha}(1,\xi)&\approx&
\cos\left(\frac{2\pi \xi}{\omega_\alpha}\right)R_{\alpha}
+\frac{4\pi^2T_D}{\omega_\alpha M}(R_{\alpha})^2m_\alpha J_0\left(\frac{4\pi t}{\omega_\alpha}\right)\cos\left(\frac{4\pi  \xi}{\omega_\alpha}\right)\nonumber\\
&&+\frac{4\pi^2T_D}{\omega_\alpha M}R_{\alpha} R_{\bar\alpha}m_{\bar\alpha}J_0\left(\frac{4\pi t}{\omega_{\bar\alpha}}\right)\cos\left(\frac{2\pi \xi}{\omega^+}\right), \label{eqn:cancel}
\end{eqnarray}

\noindent where if $\alpha=1,2$ then $\bar\alpha=2,1$, and that for $k=2$:
\begin{equation}
\cos\left(\frac{4\pi \xi^*}{\omega_\alpha}\right)R_{\alpha}(2,\xi)\approx
\cos\left(\frac{4\pi \xi}{\omega_\alpha}\right)R_{\alpha}^2. \label{eqn:cancel2}
\end{equation}

\noindent  The conductivity in this approximation is given by (after the substitution of (\ref{eqn:Deltasc}) into (\ref{eqn:startpnt})):
\begin{eqnarray}
\sigma_{zz}&=&\sum_\alpha\sigma_\alpha\int {d\xi} [-n_F'(\xi)]\left[1-\frac{\omega_\alpha}{\pi t}J_1\left(\frac{4\pi t}{\omega_{\alpha}}\right)\cos\left(\frac{2\pi\xi^*}{\omega_\alpha}\right)R_{\alpha}(1,\xi)\right.\nonumber\\
&&\left.+\frac{\omega_\alpha}{2\pi t}J_1\left(\frac{8\pi t}{\omega_{\alpha}}\right)\cos\left(\frac{4\pi\xi^*}{\omega_\alpha}\right)R_{\alpha}(2,\xi)\right]\nonumber\\
&&\times \left[1+2M^{-1}\sum_{\alpha'}R_{\alpha'}(1,\xi)m_{\alpha'}J_0\left(\frac{4\pi t}{\omega_{\alpha'}}\right)\cos\left(\frac{2\pi\xi^*}{\omega_{\alpha'}}\right)\right.\nonumber\\
&&-2M^{-1}\sum_{\alpha'}R_{\alpha'}(2,\xi)m_{\alpha'}J_0\left(\frac{8\pi t}{\omega_{\alpha'}}\right)\cos\left(\frac{4\pi\xi^*}{\omega_{\alpha'}}\right)\nonumber\\
&&\left.+\left(2M^{-1}\sum_{\alpha'}R_{\alpha'}(1,\xi)m_{\alpha'}J_0\left(\frac{4\pi t}{\omega_{\alpha'}}\right)\cos\left(\frac{2\pi\xi^*}{\omega_{\alpha'}}\right)\right)^2\right]\nonumber\\
&&+\frac{2\pi^2T_D}{\omega_\alpha}\left[-J_1\left(\frac{4\pi t}{\omega_{\alpha}}\right)\cos\left(\frac{2\pi\xi^*}{\omega_\alpha}\right)R_{\alpha}(1,\xi)\right.\nonumber\\
&&\left.+J_1\left(\frac{8\pi t}{\omega_{\alpha}}\right)\cos\left(\frac{4\pi\xi^*}{\omega_\alpha}\right)R_{\alpha}(2,\xi)\right],\label{eqn:halfway}
\end{eqnarray}

\noindent where $\sigma_\alpha =(e^2d t^2 m_\alpha)/(\pi^2 T_D)$.

Substituting (\ref{eqn:R1sub}),(\ref{eqn:cancel}) and (\ref{eqn:cancel2}) into (\ref{eqn:halfway}), integrating over $\xi$, replacing the Bessel functions with their asymptotics and gathering all the terms together, we finally obtain:
\begin{eqnarray}
\sigma_{zz}&=&\sigma\left[1+\sum_\alpha \left(\mathcal{D}_1^\alpha\cos\left(\frac{2\pi\mu}{\omega_\alpha}\right)\Theta\left(\frac{2\pi^2T}{\omega_\alpha}\right)+\mathcal{D}_2^\alpha\cos\left(\frac{4\pi\mu}{\omega_\alpha}\right)\Theta\left(\frac{4\pi^2T}{\omega_\alpha}\right)+\mathcal{D}_S^\alpha\right)\right.\nonumber\\
&&\left. +\mathcal{D}_{12}^+\cos\left(\frac{2\pi\mu}{\omega_+}\right)\Theta\left(\frac{2\pi^2 T}{\omega_+}\right)+\mathcal{D}_{12}^-\cos\left(\frac{2\pi\mu}{\omega_-}\right)\Theta\left(\frac{2\pi^2 T}{\omega_-}\right)\right].\label{eqn:scattmix}
\end{eqnarray}

Here, the unmixed amplitudes are:
\begin{equation}
\mathcal{D}_1^\alpha=2\frac{m^{1/2}_\alpha}{M}\sqrt{\frac{eB}{2\pi^2t}(1+(a_{1,\,\alpha})^2)}R_{\alpha}\cos\left(\frac{4\pi t}{\omega_{\alpha}}-\frac{\pi}{4}+\phi_{1,\,\alpha}\right),\label{eqn:IstN2}
\end{equation}
\begin{eqnarray}
\mathcal{D}_2^\alpha&=&\frac{(R_{\alpha})^2m_\alpha}{M}\left[\frac{4m_\alpha T_D}{Mt}\sqrt{1+(a_{1,\,\alpha})^2}\cos\left(\frac{4\pi t}{\omega_{\alpha}}-\frac{\pi}{4}\right)\cos\left(\frac{4\pi t}{\omega_{\alpha}}-\frac{\pi}{4}+\phi_{1,\,\alpha}\right)\right.\nonumber\\
&&\left.+2\sqrt{\frac{\omega_\alpha}{4\pi^2t}(1+(a_{2,\,\alpha})^2)}\cos\left(\frac{8\pi t}{\omega_{\alpha}}-\frac{\pi}{4}+\phi_{2,\,\alpha}\right)\right]+\mathcal{D}_S^\alpha,
\end{eqnarray}

\noindent and
\begin{equation}
\mathcal{D}_S^\alpha=\frac{eB(R_{\alpha})^2 m_\alpha}{2 \pi^2M^2 t}\left[\sqrt{1+\left(\frac{\omega_\alpha}{2 \pi t}\right)^2}\cos\left(2\left[\frac{4\pi t}{\omega_{\alpha}}-\frac{\pi}{4}+\phi_{S\alpha}\right]\right)+1\right].
\end{equation}

As in the case of oscillating $\mu$, we observe mixed frequencies as well:
\begin{eqnarray}
\mathcal{D}^{+}_{12}&=&\frac{eB R_{1}R_{2} \sqrt{m_1 m_{2}} }{M^2 t\pi^2}\left(1+\frac{2\pi^2 T_D}{\omega_+}\right)\nonumber\\
&&\times \left[\sqrt{1+(y^-)^2}\cos\left(\frac{4\pi t}{\omega_-} +\phi_{y^-}\right)+\sqrt{1+(y^+)^2}\sin\left(\frac{4\pi t}{\omega_+} +\phi_{y^+}\right)\right]
\end{eqnarray}

\noindent where
\begin{equation}
y^\pm=\left(\frac{(\omega_2 \pm \omega_1)}{4\pi t}+2\pi^2T_D\left[\frac{a_{1,\,2}}{\omega_2}\pm\frac{a_{1,\,1}}{\omega_1}\right]\right)\left(1+\frac{2\pi^2 T_D}{\omega_+}\right)^{-1},\,\phi_{y^\pm}=\arctan(y^\pm),
\end{equation}

\noindent and
\begin{equation}
\mathcal{D}^-_{12}=\frac{eB R_{1}R_{2} \sqrt{m_1 m_{2}} }{M^2 t\pi^2}
\left[\sqrt{1+(q^-)^2}\cos\left(\frac{4\pi t}{\omega_-} +\phi_{q^-}\right)+\sqrt{1+(q^+)^2}\sin\left(\frac{4\pi t}{\omega_+} +\phi_{q^+}\right)\right]
\end{equation}

\noindent where
\begin{equation}
q^\pm=\frac{(\omega_2\pm\omega_1)}{4\pi t},\,\phi_{q^\pm}=\arctan(q^\pm).
\end{equation}

\begin{figure}[htb]
\begin{center}
\epsfig{file=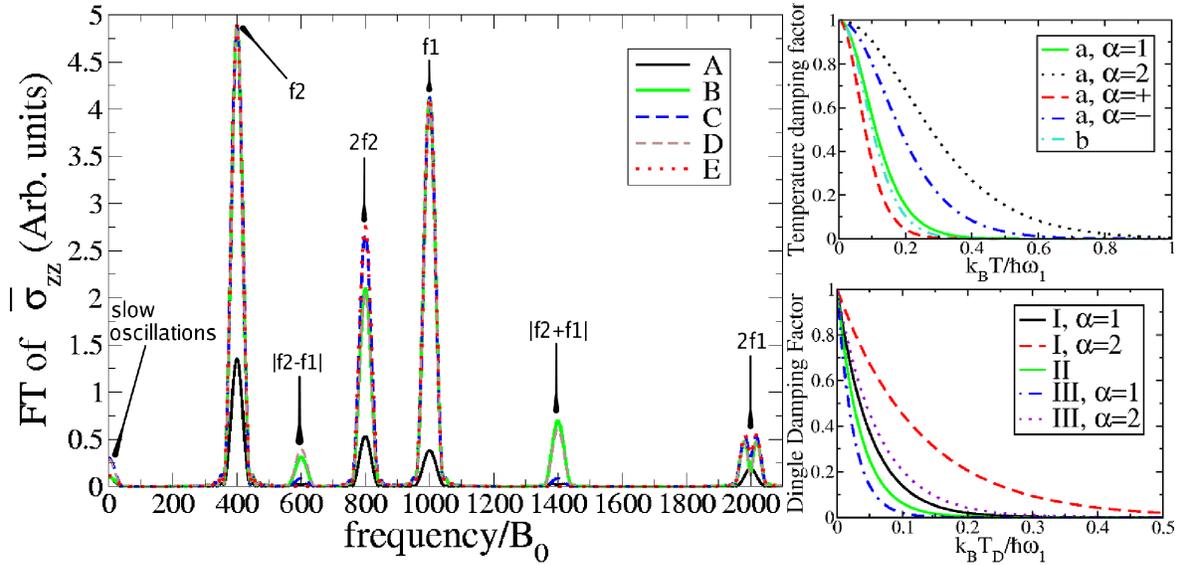, height=7.5cm}
\end{center}
\vspace{-.5cm}
\caption{\small A graph showing the Fourier transform of the SdH oscillations from $.9\le B_0/B\le.95$  with $(\omega_1/\omega_2)=(m_2/m_1)=.4$, alongside plots of how the damping factors evolve as their respective temperatures are increased. Due to the small size of the window, the slow oscillations are poorly resolved.The y axis has been rescaled by a factor of 100.  Except as otherwise noted, all parameters and legends are identical to those in Figure \ref{fig:.9}.}
\label{fig:.4}
\end{figure}

As before, the values of $a_{r,\,\alpha}$ and $\phi_{r,\,\alpha}$ and $\phi_{S\alpha}$ are given by (\ref{eqn:adef}) and (\ref{eqn:angles}) respectively.


\subsection{Canonical Ensemble: oscillating $\mu$, inter-band scattering}
\label{sec:CPcond}

Let us now treat the above system as though it were closed, and so allow $\mu$ to oscillate.  The only terms of interest to us will come from the expansion of the terms proportional to $\mathcal{D}_1^\alpha$.  From examination of (\ref{eqn:Ist1}) and (\ref{eqn:IstN2}) we can see that
\begin{equation}
\mathcal{D}_1^\alpha=\frac{m_\alpha}{M}D_1^\alpha,
\end{equation}

\noindent and it follows that the amplitudes due to the chemical potential oscillations are:
\begin{equation}
\bar\mathcal{C}_{\alpha\alpha}=\frac{m_\alpha}{M}\mathcal{C}_{\alpha\alpha},\,\bar\mathcal{C}_{\alpha\alpha'}=\frac{m_\alpha}{M}\mathcal{C}_{\alpha\alpha'},
\end{equation}

\noindent and that the expression for the conductivity is:
\begin{eqnarray}
\sigma_{zz}&=&\sigma\left[1+\sum_\alpha \left(\mathcal{D}_1^\alpha\cos\left(\frac{2\pi\mu}{\omega_\alpha}\right)\Theta\left(\frac{2\pi^2T}{\omega_\alpha}\right)\right.\right.\nonumber\\
&&\left.+\left[\mathcal{D}_2^\alpha \Theta\left(\frac{4\pi^2T}{\omega_\alpha}\right)-\bar\mathcal{C}_{\alpha\alpha}\Theta\left(\frac{2\pi^2T}{\omega_\alpha}\right)^2\right]\cos\left(\frac{4\pi\mu}{\omega_\alpha}\right)+\mathcal{D}_S^\alpha+\bar\mathcal{C}_{\alpha\alpha}\Theta\left(\frac{2\pi^2T}{\omega_\alpha}\right)^2\right)\nonumber\\
&&+\left[\mathcal{D}_{12}^+\Theta\left(\frac{2\pi^2 T}{\omega_+}\right)-(\bar\mathcal{C}_{12}+\bar\mathcal{C}_{21})\Theta\left(\frac{2\pi^2 T}{\omega_1}\right)\Theta\left(\frac{2\pi^2 T}{\omega_2}\right)\right]\cos\left(\frac{2\pi\mu}{\omega_+}\right)\nonumber\\
&&\left.+\left[\mathcal{D}_{12}^-\Theta\left(\frac{2\pi^2 T}{\omega_-}\right)+(\bar\mathcal{C}_{12}+\bar\mathcal{C}_{21})\Theta\left(\frac{2\pi^2 T}{\omega_1}\right)\Theta\left(\frac{2\pi^2 T}{\omega_2}\right)\right]\cos\left(\frac{2\pi\mu}{\omega_-}\right)\right].
\end{eqnarray}

In this case, all the second order terms are modified by the chemical potential oscillations.

\section{Discussion}

\begin{figure}[t]
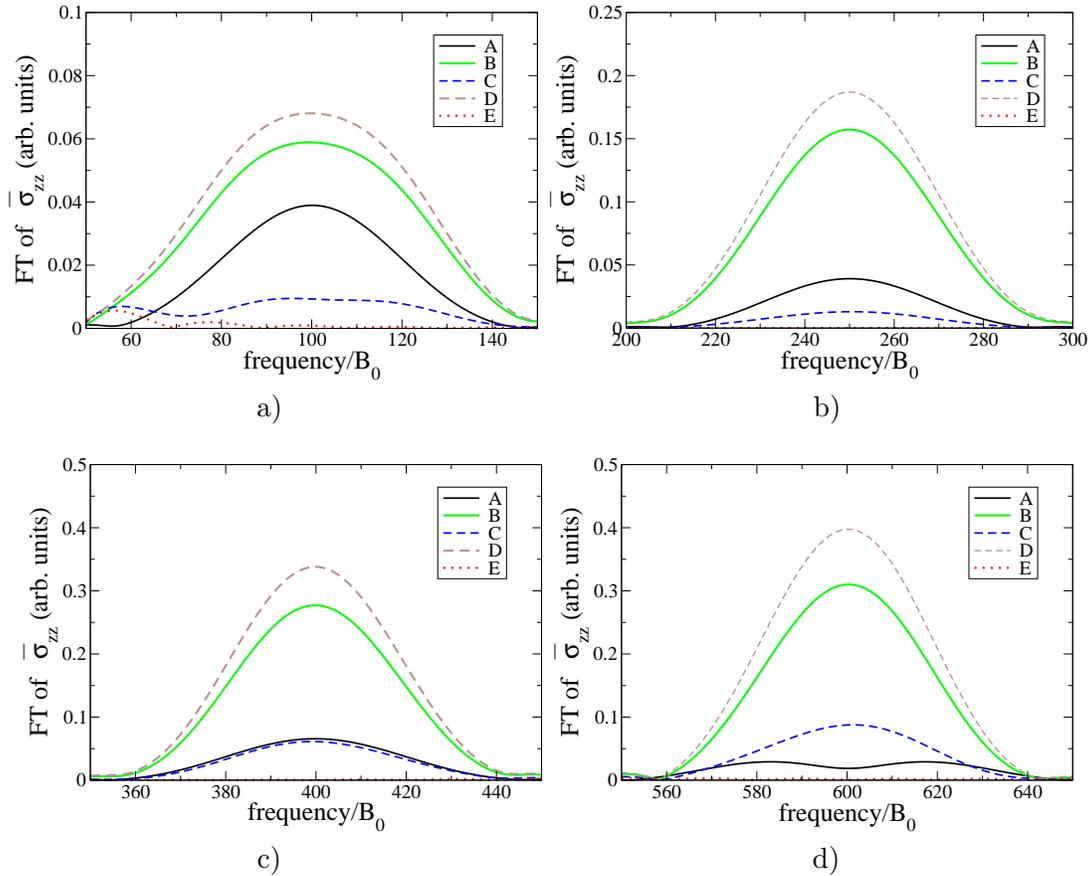

\vspace{0.5cm}
\begin{center}
\epsfig{file=fminusr.9.eps, height=5cm}
\epsfig{file=fminusr.75.eps,height=5cm}
\end{center}
\begin{center}
\vspace{-.75cm}
a) \hspace{6.8cm} b)
\end{center}
\vspace{-.5cm}
\begin{center}
\epsfig{file=fminusr.6.eps, height=5cm}
\epsfig{file=fminusr.4.eps,height=5cm}
\end{center}
\begin{center}
\vspace{-.75cm}
c) \hspace{6.8cm} d)
\end{center}
\vspace{-.5cm}
\caption{\small Figures showing the details of the $|f_2-f_1|$ amplitudes of $(m_2/m_1)$ values of a) 0.9, b) 0.75, c) 0.6 and d) 0.4 .}\label{fig:fmzoom}
\end{figure}

Figures~\ref{fig:.9} to \ref{fig:.4} display for purposes of illustration the Fourier transforms of the oscillatory components $\bar\sigma_{zz}$ of $(\sigma_{zz}-\sigma)/\sigma$ given by the expressions in the previous sections.  The inclusion of plots of the behaviour of the Dingle and temperature reduction factors in the figures should facilitate the extrapolation the results given here to regimes where $R_\alpha\ll1$, where our analytic calculations are more valid; however, it would be suprising if the qualitative elements of our results were not preserved even at values of $R_\alpha$ near to unity.  For the purposes of reference to a real material, Cole {\em et al.}'s \cite{Coleetc} measurements of the effective mass in the two-band system of GaAs--(Ga,Al)As heterojunctions indicate a  ratio of masses somewhere in the region $.3\lapprox (m_2/m_1)\lapprox .45$.

Note that in order to clean up the signal and remove spurious `ringing' due to the finite window size, the data was passed through a Hanning window prior to the operation of the numerical Fourier transform (see, for example, reference \cite{Stade}).

\begin{figure}[t]
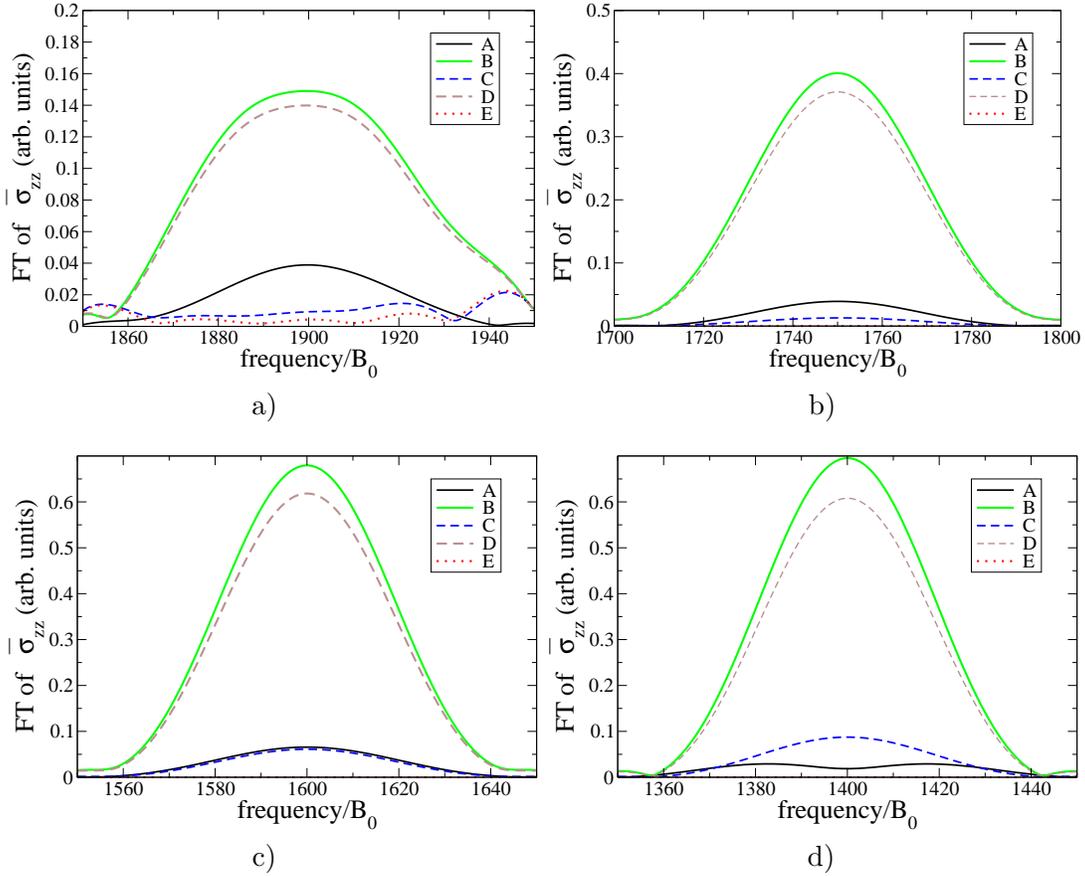

\vspace{0.5cm}
\begin{center}
\epsfig{file=fplusr.9.eps, height=5cm}
\epsfig{file=fplusr.75.eps,height=5cm}
\end{center}
\begin{center}
\vspace{-.75cm}
a) \hspace{6.8cm} b)
\end{center}
\vspace{-.5cm}
\begin{center}
\epsfig{file=fplusr.6.eps, height=5cm}
\epsfig{file=fplusr.4.eps,height=5cm}
\end{center}
\begin{center}
\vspace{-.75cm}
c) \hspace{6.8cm} d)
\end{center}
\vspace{-.5cm}
\caption{\small Figures showing the details of the $|f_2+f_1|$ amplitudes of $(m_2/m_1)$ values of a) 0.9, b) 0.75, c) 0.6 and d) 0.4 .}\label{fig:fpzoom}
\end{figure}
Details of the mixing frequencies are shown in Figures~\ref{fig:fmzoom} and \ref{fig:fpzoom}.  Perhaps the most obvious and interesting feature is that the signature of the two kinds of mixing is not the same: the amplitudes of the $|f_2 +f_1|$  and the $|f_2 -f_1|$ mixing are identical in the cases where the mixing is due to only oscillations in $\mu$, but in cases where  mixing through inter-band scattering is present, the $|f_2 -f_1|$ mixing amplitude is less than that of the $|f_2 +f_1|$ mixing (the exception is case C at a ratio of $.9$, most likely due to the small size of its amplitude and its being interfered with by neighboring peaks).  In general, the amplitude of the mixing peaks grows more pronounced as the two bands become less similar (that is, as $m_2$ becomes smaller than $m_1$), at least for these values of $m_2$, apart from case A at a ratio of .4, where a splitting of the peaks reduces their size. It should also be noted from the structure of the LK factors in the mixing terms of (\ref{eqn:tmuosc}) and (\ref{eqn:scattmix}) that the amplitudes behave differently with respect to the temperature, and that this provides a further method of distinguishing between the two sources of oscillation (see \cite{FortinZiman} for the relevence of this point in the context of the theory of magnetic breakdown).

\begin{figure}[t]
\vspace{0.2cm}
\begin{center}
\epsfig{file=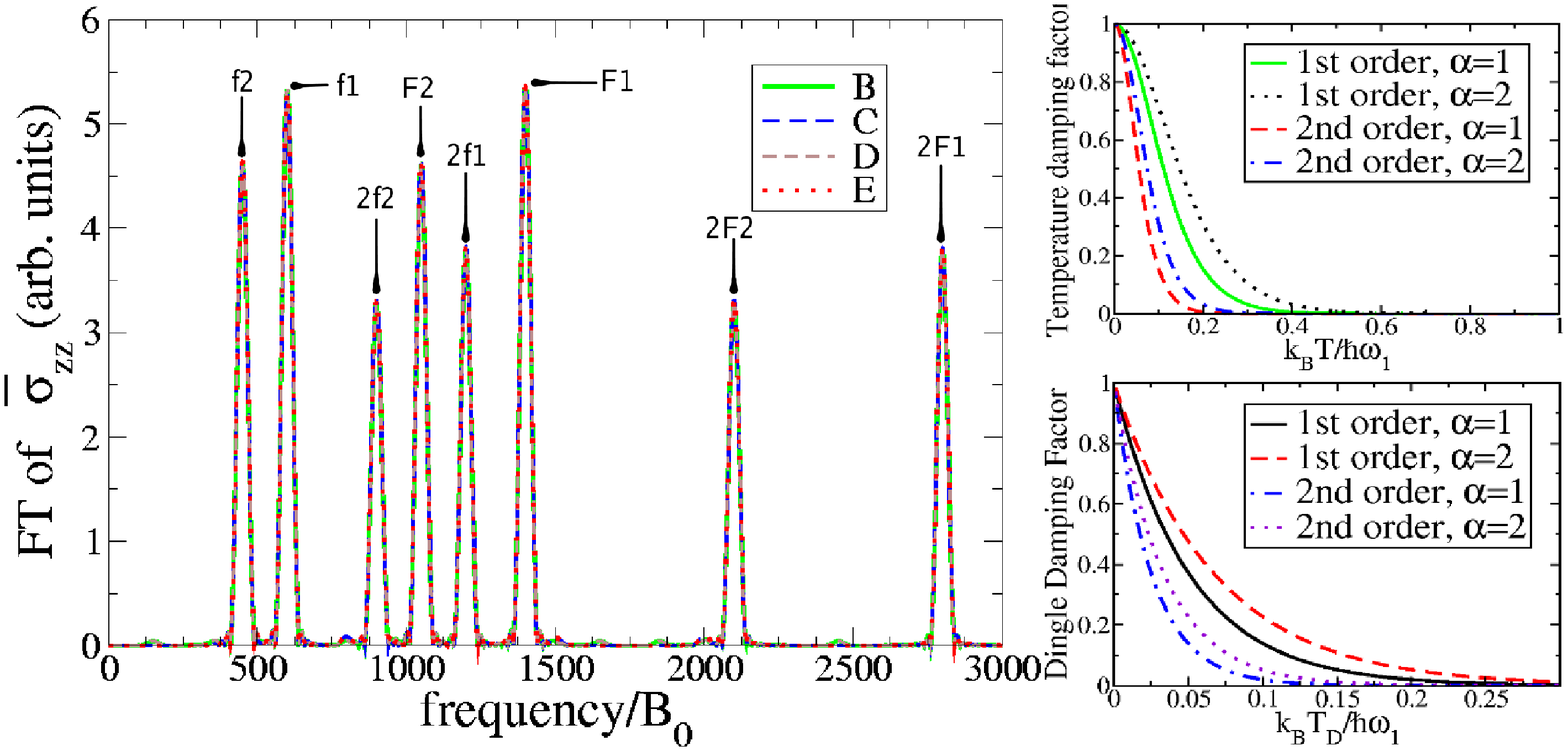, height=7.5cm}
\\\vspace{-.5cm}
\end{center}
\begin{center}
\epsfig{file=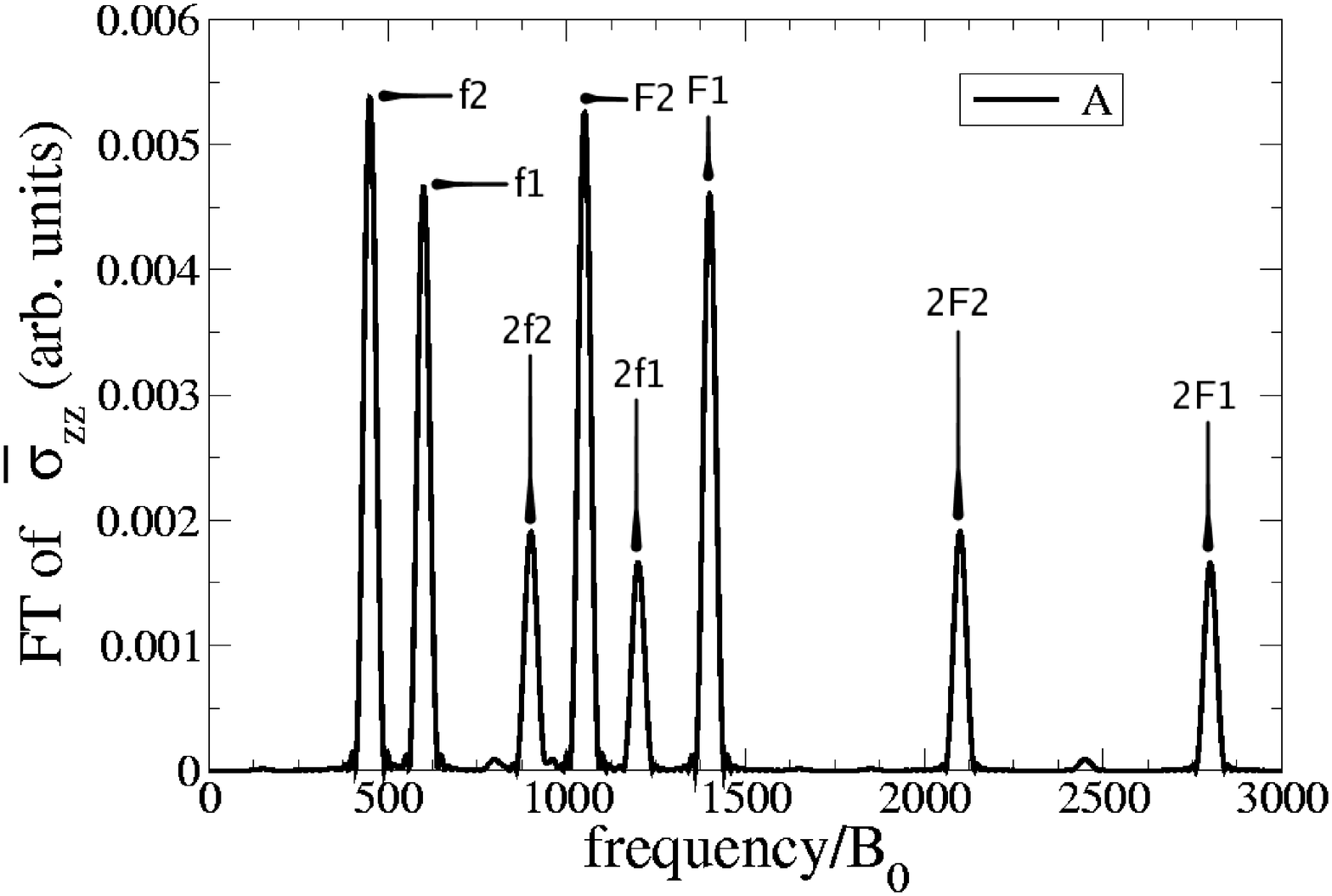,height=7.5cm}
\\\vspace{-1cm}
\end{center}
\caption{\small Fourier transforms of the SdH oscillations in the large $t$ limit for a window of $.9\le B_0/B\le.95$ where the frequency is given in terms of $B/B_0$ and $\mu=5t$,  $(\omega_{1}/\omega_{2})=(m_2/m_1)=.75$,$k_ B T=2.5\times10^{-7}t$, $B_0=m_1t/200\hbar e$, which is around 4 Tesla when $m_1=m_e$ and $t=.1$eV , and we have set $T_D$ to be zero. The cosine and sine functions in the amplitudes have split the peaks of each harmonic in two; these are labeled $f$ and $F$ corresponding to the `neck' and `belly' frequencies repectively.  The y axis in both graphs has been rescaled by a factor of 1000; note that the amplitudes in the case of oscillating $\mu$ alone (A) are {\em much} smaller than those of the other cases.  The legend in the Fourier Transform plots  should be interpreted as follows: A -- closed system, fixed $\tau$; B -- open system, inter-band scattering; C -- closed system, intra-band scattering; D -- closed system, interband scattering; E -- open system, intraband scattering.  The plots of the damping factors show how the first and second order damping terms vary with their respective temperatures. }\label{fig:3dlim}
\end{figure}

Our calculations can also be applied to quasi-three dimensional metals, where $t$ is less than or comparable to $\mu$. Figure~\ref{fig:3dlim} shows the Fourier transform of $\bar\sigma_{zz}$ when $\mu=5t$,  $(\omega_{1}/\omega_{2})=(m_2/m_1)=.75$,$k_ B T=2.5\times10^{-7}t$, $B_0=m_1t/(200\hbar e)$, and we have set $T_D$ to be zero, in order to clarify the harmonic behavior of the amplitudes.  In this case our results correspond to a Fourier series truncated after the second harmonics of each band, and we can see that the mixing effects due to scattering or chemical potential oscillations are suppressed in this limit.  We can observe that the ratio between the first and second harmonics is $\approx 2^{1/2}$ in the cases where we have scattering, and $\approx 2^{3/2}$ in the case of oscillating $\mu$ alone -- this difference is due to our neglecting the contribution from scattering to the oscillations in the latter case, which, as we can see from a comparison of the amplitudes in the graphs, is the dominant source of oscillations in three dimensions (as one might expect from the results of calculations in 3D metals \cite{Kubo2}).  Naturally, at finite temperatures, the amplitudes will decay correspondingly more quickly due to the effect of the temperature and Dingle damping factors, which can be extrapolated from their respective plots displayed in the figure.

\section{Conclusion}

We have examined possible sources of frequency mixing in the SdH oscillations of  multiband quasi-2D metals in the canonical and grand canonical metallic ensembles in the intermediate values of the field where $4\pi t>\omega_\alpha$.  We considered a closed system with a fixed relaxation time $\tau$, an open system with oscillations in $\tau$ arising from intra-band scattering, a closed system with oscillations in $\tau$ arising from intra-band scattering, an open system with oscillations in $\tau$ arising from inter-band scattering, and lastly a closed system with inter-band scattering with oscillations in $\tau$ arising from inter-band scattering.  

In all cases other than the second we discover some measure of frequency mixing.  However, the behavior of the mixing amplitudes are slightly different for each case where they occured, which may allow the mechanisms involved in experimental systems to be distingished. In the fourth and fifth cases, we observe frequency mixing due to scattering, which is interesting as we would expect such an effect in the dHvA oscillations to be virtually negligible (as argued in Section \ref{sec:calc}).

\section{Acknowledgments}

The authors would like to thank the EPSRC for funding this work (grant No.~EP/D035589).

\end{document}